\documentclass[%
 reprint,
 amsmath,amssymb,
 aps,
pra,
floatfix,
]{revtex4-1}
\usepackage{lipsum}
\usepackage{mathrsfs}
\usepackage{graphicx,lipsum}
\usepackage{dcolumn}
\usepackage{bm}
\usepackage[dvipsnames]{xcolor}
\usepackage{array,multirow}
\usepackage{bm}
\usepackage{bbm}
\usepackage{amsmath}
\usepackage{epstopdf}

\usepackage{appendix}
\usepackage{xcolor}
\usepackage{graphicx}
\usepackage{dcolumn}

\begin{document}

\preprint{APS/123-QED}

\title{Phase covariant channel:  Quantum speed limit of evolution}

\author{Riya Baruah}
\altaffiliation[baruah.1@iitj.ac.in]{}
\author{K.G. Paulson}%
 \email{paulsonkgeorg@gmail.com}


\author{Subhashish Banerjee}
\email{subhashish@iitj.ac.in}
\affiliation{Indian Institute of Technology, Jodhpur-342030, India}


\date{\today}

\begin{abstract}
The quantum speed of evolution for the phase covariant map is investigated. This involves absorption, emission and dephasing processes. We consider the maps under various combinations of the above processes to investigate the effect of phase covariant maps on quantum speed limit time. For absorption-free phase covariant maps, combinations of dissipative and  CP-(in)divisible  (non)-Markovian dephasing  noises are considered.  The role of coherence-mixedness balance on the speed limit time is checked in the presence of both vacuum and finite temperature effects. We also investigate the rate at which Holevo's information changes and  the action quantum speed of evolution for specific cases of the phase covariant map.



\end{abstract}

\maketitle

\section{\label{sec:level1}Introduction}
Nowadays, it is an established fact that one can manipulate the impact of quantum noise on quantum systems productively~\cite{sbbook}. A good amount of work has been devoted to investigating the memory effects of noise on the dynamics of quantum systems~\cite{rhp,hall,basano,pradeep,nmdshrikant,sbarko}. In general, (non)-Markovianity discusses the nature of the system's correlations with the environment. Quantum coherence and correlations are significant resources for quantum technology~\cite{Paulson2014,adesso2016measures,streltsov2017colloquium,paulson2019tripartite,sapienza2019correlations}. Non-Markovianity influences the quantum resources in both beneficial and unfavourable ways; consequently, the investigation of quantum correlation and coherence becomes highly significant~\cite{nmcrypto,nmpingpong,samyasb}. Along with the memory effects on quantum resources, it is pertinent to discuss the evolution speed of quantum systems. It's been shown that energy-time uncertainty reveals the bound on the speed of the evolution of quantum states~\cite{mandelstam1945}. Initially, speed limit time was derived for the dynamics between the orthogonal states for isolated systems~\cite{margolus1998}. Later, quantum speed limit (QSL) time for the time-independent systems was extended to the arbitrary quantum states~\cite{giovannetti2003quantum}. Further, the speed limit for the evolution between the states with the arbitrary angle for a driven quantum system has been determined~\cite{deffner2013energy}. Recently, the speed of evolution between arbitrary states for open quantum systems~\cite{del2013quantum,deffner2013quantum,davidovich} has become a lively research topic and is the central theme of the present work.\newline

Not only from the dynamical perspective but also the system's characteristics revealed by the limit on the speed of evolution display its distinguishable role in quantum communication and technology. To list a few, the bound on speed limit time reveals how fast the quantum information can be communicated, the maximum rate at which information can be processed, and the precision limit in quantum metrology~\cite{bekenstein1981energy,lloyd2000ultimate,giovannetti2011advances} among others. Even though there exists no direct connection between non-Markovianity, a class of which is identified by information backflow and quantum speed limit time ($\tau_{QSL}$)~\cite{teittinen2019there}, it has been shown that $\tau_{QSL}$ could be realized as a witness of the decay-revival mechanism of quantum correlations~\cite{paulson2021hierarchy} for a certain class of quantum noises. In~\cite{deffner2013quantum,xu2014quantum,mirkin2016quantum,paulson2021effect}, it has been seen that quantum non-Markovianity may speed up the evolution of quantum states.
It is also known that non-Markovianity is not always required to speed up quantum evolution~\cite{cai2017quantum,marian2021quantum,lan2022geometric}. From the practical point of view, $\tau_{QSL}$ finds many applications in a wide range of fields~\cite{2017}. \newline
The phase-covariant map describes the physical processes involving absorption, emission and pure dephasing. This provides a convenient platform to study both (non)-unital processes from a common perspective.
In the present work, we estimate the $\tau_{QSL}$ for single-qubit states evolving under the phase covariant channel~\cite{Smirne_2016}.  
This channel can be thought of as an approximation of the general spin-Boson problem ~\cite{Hasse}. Work on similar lines was initiated recently in ~\cite{Sabrina21}. Here, in addition to the impact of various processes like heating, dissipation and dephasing on ${\tau_{QSL}}$, we also consider the role of coherence and mixing,  as well as the purity of initial states. Coherence is one of the central features of quantum physics ~\cite{plenio}. Further, an open system evolution generally makes a system's states mixed. Hence, it is meaningful to ask how the balance between these two processes, {\it viz.} coherence and mixing, impacts the dynamics ~\cite{samyasb,khushboosb}. We further consider different combinations of CP-(in)divisible (non)-Markovian quantum channels and a phenomenological model and show how these combinations influence the speed  of quantum evolution for both pure and mixed initial states. The influence of thermal bath on $\tau_{QSL}$, along with the rate at which the upper bound for Holevo's information changes, are also checked.\newline
The present work is structured as follows. Section II discusses the prerequisites for the current work, which contains the details of the phase covariant channel, quantum speed limit, and the measure of non-Markovianity used here. In Sec. III, we investigate $\tau_{QSL}$ and the impact of coherence-mixedness trade-off on $\tau_{QSL}$ for different quantum phase covariant noises for an initial pure state. Quantum speed limit time for an initially mixed state is discussed in Sec. IV. In Sec. V, we briefly discuss the recently introduced action quantum speed limit for the phase covariant map, followed by the concluding remarks in Sec. VI.
\section{\label{sec:level2}Preliminaries}
Here we present the preliminary information required for this work. This begins with a brief overview of the phase covariant map, followed by a discussion of the QSL time using the geometric approach. Since the interplay of coherence and mixing, along with the non-Markovian behaviour of the combination of different quantum channels, is central to the present work, these notions are  briefly introduced.
\subsection{\label{sec:2.a}The phase covariant map}
The master equation for a single qubit phase covariant dynamics has the form ~\cite{Smirne_2016}:
\begin{align}
\begin{split}\label{eq:phase_cov}
    \frac{d\rho(t)}{dt}&=-i\frac{\omega(t)}{2}[\sigma_z,\rho(t)]+\frac{\gamma_1(t)}{2}\mathcal{L}_1(\rho(t))+\frac{\gamma_2(t)}{2}\mathcal{L}_2(\rho(t))\\
         & \quad  +\frac{\gamma_3(t)}{2}\mathcal{L}_3(\rho(t)),
\end{split}
\end{align}
where 
\begin{align}
\mathcal{L}_1(\rho(t))&=\sigma_+\rho(t)\sigma_--\frac{1}{2}\{\sigma_-\sigma_+,\rho(t)\}, \label{eq:3} \\
\mathcal{L}_2(\rho(t))&=\sigma_-\rho(t)\sigma_+-\frac{1}{2}\{\sigma_+\sigma_-,\rho(t)\}, \label{eq:4} \\
\mathcal{L}_3(\rho(t))&=\sigma_z\rho(t)\sigma_z-\rho(t),
\end{align}
and ${\sigma_\pm=(\sigma_x\pm i \sigma_y)/2}$.
The action of the phase covariant map on an arbitrary single-qubit density matrix ${\rho(0)}$ is ~\cite{teittinen2018revealing}:
\begin{equation} \label{eq:den}
    \Phi_t(\rho(0))=\rho(t)=
    \begin{pmatrix}
    1-p_1(t) & \alpha^*(t) \\
    \alpha(t) & p_1(t)
    \end{pmatrix},
\end{equation}
where
\begin{align}
    p_1(t)&=e^{-\Gamma(t)}[G(t)+p_1(0)],\\
    \alpha(t)&=\alpha(0)e^{i\Omega(t)-\Gamma(t)/2-\widetilde{\Gamma}(t)},
\end{align}
and 
\begin{align}\label{eqn:den_mat}
    \Gamma(t)&=\int_{0}^{t} \frac{\gamma_1(\tau)+\gamma_2(\tau)}{2} d\tau, \\
    G(t)&=\int_{0}^{t} e^{\Gamma(\tau)} \frac{\gamma_2(\tau)}{2} d\tau, \\
    \Omega(t)&=\int_{0}^{t}2\omega(\tau)d\tau, \\
    \widetilde{\Gamma}(t)&=\int_0^t \gamma_3(\tau)d\tau.
\end{align}
Here, ${\gamma_1(t)}$,~${\gamma_2(t)}$,~${\gamma_3(t)}$ correspond to energy gain, energy loss and pure dephasing rates, respectively. Also, ${\Omega(t)}$ corresponds to rotations around the $z$-axis of the Bloch ball. Phase covariant dynamics for a single qubit satisfies the relation ${\text{exp}(-i\sigma_z\phi)\Phi[\rho]\text{exp}(i\sigma_z\phi)=\Phi[\text{exp}(-i\sigma_z\phi)\rho\text{exp}(i\sigma_z\phi)]}$ for all real ${\phi}$~\cite{RussianFillipov}. It can be shown that phase covariant dynamics implies uniform deformation of the $x$ and $y$ Bloch vectors and permits a deformation as well as translation of the $z$ Bloch vector.

\subsection{\label{sec:2.b}Quantum speed limit time}
Mandelstam and Tamm (MT)~ and Margolus and Levitin (ML)-type bounds on speed limit time are estimated using the geometric approach to quantify the closeness between the initial and final states. Here, the Bures angle measures the distance between two quantum states. In~\cite{deffner2013quantum}, for the initial pure state $\rho_0=\vert\psi_0\rangle\langle\psi_0\vert$, a bound on the speed limit time based on Bures angle $\mathcal{B}(\rho_0,\rho_t)$ is,
\begin{equation}
    \tau_{QSL}=\max\Bigg\{\frac{1}{\Lambda^{\textrm{op}}_{\tau}},\frac{1}{\Lambda^{\textrm{tr}}_{\tau}},\frac{1}{\Lambda^{\textrm{hs}}_{\tau}}\Bigg\} \sin^2[\mathcal{B}],
    \label{B_spdlmt}
\end{equation}
where  $\mathcal{B}(\rho_0,\rho_t)=\arccos(\sqrt{\langle\psi_0\vert\rho_t\vert\psi_0\rangle})$.
\begin{equation}
    \Lambda^{\textrm{op,tr,hs}}_{\tau}=\frac{1}{\tau}\int^{\tau}_{0} dt \vert\vert \mathcal{L}(\rho_t)\vert\vert_{\textrm{op,tr,hs}},
    \label{B_spdlmt_1}
\end{equation}
where  ${\Lambda^{\textrm{op}}_{\tau}}$,${\Lambda^{\textrm{tr}}_{\tau}}$, and ${\Lambda^{\textrm{hs}}_{\tau}}$ are the operator, Hilbert-Schmidt and trace norms, respectively. Operators satisfy the von Neumann trace inequality 
$\vert\vert A\vert\vert_{\textrm{op}}\leq\vert\vert A\vert\vert_{\textrm{hs}}\leq\vert\vert A\vert\vert_{\textrm{tr}}$, which gives, $1/\Lambda^{\textrm{op}}_{\tau}\geq1/\Lambda^{\textrm{hs}}_{\tau}\geq1/\Lambda^{\textrm{tr}}_{\tau}$.  The tighter bound on the quantum speed limit time is achieved by using operator norm of the generator. An upper bound on fidelity for any density matrices ${\rho_1}$ and ${\rho_2}$  shows that ~\cite{miszczak2008sub} $\mathcal{F}(\rho_1,\rho_2)\leq \textrm{tr}\rho_1\rho_2+\sqrt{(1-\textrm{tr}\rho_1^2)(1-\textrm{tr}\rho_2^2)}$. Making use of this super-fidelity, a  bound on quantum speed limit time for both pure and mixed initial states can be written by multiplying the RHS in (Eq.~\ref{B_spdlmt_1}) by a factor $\Bigg(1+\sqrt{\frac{1-\textrm{tr}\rho_0^2}{1-\textrm{tr}\rho_t^2}}\Bigg)$~\cite{wu2020quantum}.

\subsection{\label{sec:2.c} Coherence-mixing balance and the upper limit of the Holevo Bound}
The Mixedness of a quantum system imposes a limit on the amount of quantum coherence it can possess~\cite{PhysRevA.91.052115,samyasb}. For a d-level system, this trade-off can be expressed as an inequality: 
\begin{equation}
    M_{cl}=\frac{C_{l_1}^2(\rho)}{(d-1)^2}+M_l(\rho)\leq 1.
\end{equation}
For a two-level system, using the density matrix equation Eq.~(\ref{eq:den}), the trade-off equation can be shown to be in the following helpful form
\begin{equation}\label{eq:trade-off}
    4p_1(t)(1-p_1(t))\leqslant 1,
\end{equation}
see the appendix for the details of the derivation. It was shown recently shown that the rate at which accessible information, quantified by the Holevo quantity ${\chi}$ changes is upper bounded by the quantum speed limit time as ~\cite{2017, acconcia2017quantum}, 
\begin{equation}
    \dot{\chi}\leqslant\frac{\Delta\chi}{\tau_{QSL}}.
\end{equation}
Here, ${\Delta\chi}$ represents the change of ${\chi}$. This suggests that ${\tau_{QSL}}$, in particular, ${1/\tau_{QSL}}$, upper bounds the rate with which the accessible information changes.

\subsection{\label{sec:2.d}Self similarity measure of non-Markovianity}
Recently, a measure, which we call the SSS measure, was defined that approaches non-Markovian behaviour from the perspective of temporal self-similarity, the property of a system dynamics wherein the propagator between two intermediate states is independent of the initial time ~\cite{SSSnat2020}. In particular, it quantifies non-Markovian behaviour in terms of deviation ${\zeta}$ from the temporal self-similarity
\begin{equation}
    \zeta=\min_{\mathcal{L*}}\frac{1}{T}\int_0^T||\mathcal{L}(t)-\mathcal{L^*}||_{\text{tr}}~dt.
    \label{tss_measure}
\end{equation}
Here ${||A||_{\text{tr}}=\text{tr}\sqrt{A^\dagger A}}$ is the trace norm of matrix A, ${\mathcal{L}}(t)$ is the Lindbladian corresponding to the time-homogeneous master equation and ${\mathcal{L^*}}$ is a time-independent Lindblad generator.

\begin{figure}[h]
\includegraphics[width=\linewidth]{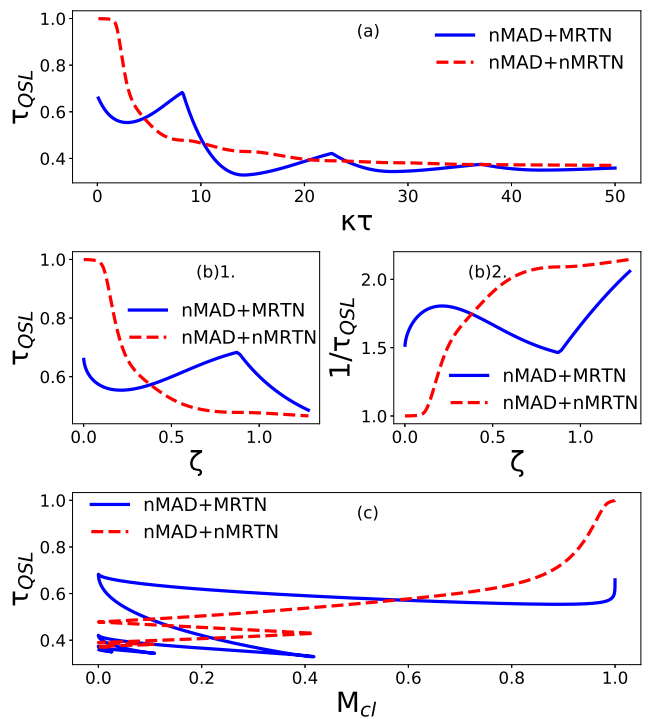}
\caption{Quantum speed limit time ${\tau_{QSL}}$ vs ${\kappa\tau}$, $\tau_{QSL}$ and ${\frac{1}{\tau_{QSL}}}$ vs ${\zeta}$, and ${\tau_{QSL}}$ as a function of ${M_{cl}}$ are shown in subplots (a), (b), and (c), respectively for different  combinations of non-Markovian amplitude damping $(l=0.1k)$ and RTN dephasing channels with ${\frac{1}{\sqrt{2}}(|0\rangle+|1\rangle)}$ as the initial state. ${\eta=\kappa}$;  ${\alpha=\kappa}$ and ${\alpha=0.1\kappa}$ for non-Markovian (nMRTN) and Markovian RTN (MRTN) channels, respectively. Actual driving time ${\tau=1}$.}\label{fig:1}
\end{figure}

\section{\label{sec:level3} Quantum speed limit time: analysis of various phase covariant maps}
The master equation for single qubit phase covariant dynamics (Eq.~\ref{eq:phase_cov}) discusses the evolution of quantum states under various physical processes such as absorption, emission and dephasing which are characterised by the rate constants $\gamma_1(t), \gamma_2(t)$ and $\gamma_3 (t)$, respectively, from the framework of phase covariant map. Here, we investigate the   quantum speed limit time for pure and mixed initial  qubit states under the combination of different (non)-unital (non)-Markovian quantum channels and analysis the impact of coherence-mixing on ${\tau_{QSL}}$.

\subsection{\label{sec:3.a}Non-Markovian amplitude damping and random telegraph noise}

Here, we consider the case of absorption-free phase covariant dynamics. A combination of the non-unital non-Markovian amplitude damping (nMAD) channel  and the unital random telegraph noise (RTN), a pure  dephasing channel ~\cite{SSSnat2020} is taken into consideration. Amplitude damping channel models the physical processes like spontaneous emission. For (non)-Markovian RTN dephasing channels, decoherence processes are induced by low-frequency noise modelled through stochastic processes.  Since it's a absorption-free process we take ${\gamma_1(t)=0}$ in Eq.~\ref{eq:phase_cov}.
For the nMAD channel, ${\gamma_2(t)=-\frac{4}{\Lambda(t)}\frac{d\Lambda(t)}{dt}}$ where ${\Lambda(t)=e^{-l t/2}(\text{cosh(zt/2)}+\frac{l}{z}\text{ sinh(zt/2)})}$~\cite{paulson2021effect} is the decoherence function. The decoherence rate then becomes
\begin{equation}\label{eq:12}
    \gamma_2(t)=\frac{4 \kappa l\text{ sinh(zt/2)}}{\text{z cosh(zt/2)}+l\text{ sinh(zt/2)}},
\end{equation}
where ${z=\sqrt{l^2-2\kappa l}}$, ${\kappa}$ describes the qubit-environment coupling strength and ${l}$ is the spectral width related to the reservoir correlation time. The dynamics are Markovian in the region ${l>2k}$, whereas it is non-Markovian in the region ${l<2k}$. 
The dephasing rate for RTN is ${\gamma_3(t)=-\frac{1}{\Lambda(t)}\frac{d\Lambda(t)}{dt}}$ where $\Lambda(t)=e^{-\eta t}(\text{cos}(\mu\eta t) + \frac{1}{\mu}\text{sin}(\mu \eta t))$ and is
\begin{equation}
    \gamma_3(t)=\frac{\eta(\mu^2+1)\text{ sin}(\mu\eta t)}{\mu\text{ cos}(\mu\eta t) + \text{sin}(\mu\eta t)},
\end{equation}
where ${\mu=\sqrt{(\frac{2\alpha}{\eta})^2-1}}$. Here, ${\eta}$ is the spectral bandwidth and ${\alpha}$ is the coupling strength between the qubit and the reservoir. Depending on whether ${(\frac{2\alpha}{\eta})^2>1}$ or ${(\frac{2\alpha}{\eta})^2<1}$, the dynamics is non-Markovian or Markovian, respectively. Using Eq.~\ref{tss_measure}, we calculate the channel's memory $\zeta$, $\zeta>0$ indicates the presence of the memory.\\
Fig.~\ref{fig:1}(a) depicts the behaviour of ${\tau_{QSL}}$ with respect to the dimensionless time (${\kappa\tau}$) for the maximally coherent initial state ${\frac{1}{\sqrt{2}}(|0\rangle+|1\rangle)}$. We have QSL time and $1/\tau_{QSL}$ vs $\zeta$  in Figs.~\ref{fig:1}(b1) and \ref{fig:1}(b2), respectively. $M_{cl}$ vs QSL time is shown in Fig.~\ref{fig:1}(c). We find that a more  non-Markovian combination of channels does not always lead to the speed-up of quantum evolution. It generally depends on the strength of the coupling and the initial state chosen. The wiggling nature of the QSL curve could be attributed to the RTN noise in the non-Markovian regime due to the highly oscillatory nature of the noise in this regime. The ${\tau_{QSL}}$ vs  ${M_{cl}}$ plot brings out the influence of coherence and mixing on the speed of evolution.
\begin{figure}[h]
\includegraphics[width=\linewidth]{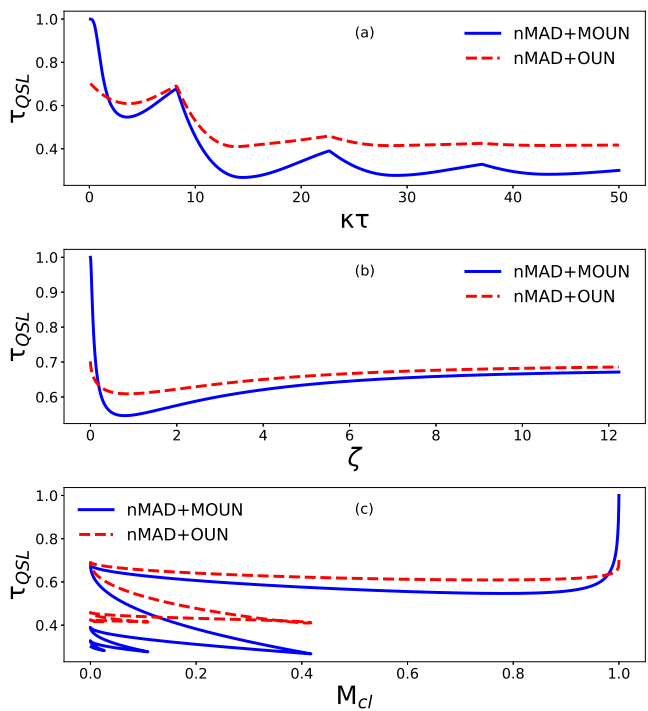}
\caption{ Quantum speed limit time ${\tau_{QSL}}$ vs ${\kappa\tau}$,  (b) $\tau_{QSL}$ vs $\zeta$  (c) ${\tau_{QSL}}$ as a function of ${M_{cl}}$ plotted for a combination of non-Markovian amplitude damping and OUN dephasing channels with ${\frac{1}{\sqrt{2}}(|0\rangle+|1\rangle)}$ as the initial state. The parameters are $ l=p=0.1\kappa, m=\kappa$ for non-Markovian amplitude damping and OUN channels. $l=0.1\kappa$ and ${p=0.1}$ for the case of MOUN.} \label{fig:2}
\end{figure}
\subsection{\label{sec:3.b}Non-Markovian amplitude damping and modified Ornstein–Uhlenbeck noise}
Now, RTN noise in the previous case (Sec.~\ref{sec:3.a}) is replaced by the Modified Ornstein-Uhlenbeck noise (OUN),~\cite{yu2010}  a stationary Gaussian random process. The spin of an electron interacting with a magnetic field subject to stochastic fluctuations is a physical scenario that leads to the occurrence of OUN. OUN is  CP divisible, but it is still non-Markovian as it exhibits memory effects, as captured by the SSS measure ~\cite{SSSnat2020}. This noise has a well-defined Markovian limit, referred to here as MOUN.
The OUN dephasing rate is ${\gamma_3(t)=-\frac{1}{\Lambda(t)}\frac{d\Lambda(t)}{dt}}$ where ${\Lambda(t)=\text{exp}(-\frac{p}{2}(t+\frac{1}{m}(e^{-m t} -1)))}$ which is calculated as
\begin{equation}
    \gamma_3(t)=\frac{p}{2}(1-e^{-m t}).
\end{equation}
Here, $p$ is the inverse of the effective relaxation time, and $m$ is related to the noise bandwidth. This channel is Markovian in the limit ${\frac{1}{m}\rightarrow\infty}$ for which case ${\Lambda(t)=e^{-p t/2}}$ and ${\gamma_3(t)=p/2}$. We use Eq.~(\ref{eq:12}) for ${\gamma_2(t)}$ and $\gamma_1(t)=0$.\\

\begin{figure}[!htb]
\includegraphics[width=\linewidth]{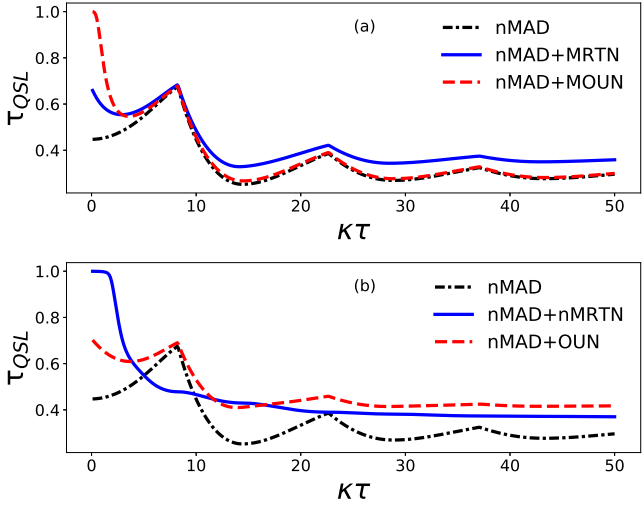}
\caption{QSL time is plotted as a function of $\kappa\tau$ for phase covariant maps with non-Markovian amplitude damping channel and  (non)-Markovian RTN and OUN channels.}
\label{fig:all_oun_rtn}
\end{figure}

This phase covariant channel exhibits memory effects and is quantified using Eq.~\ref{tss_measure}. The variation of ${\tau_{QSL}}$ for OUN in combination with nMAD to coupling strength is depicted in Fig.~\ref{fig:2}(a) for the initial maximally coherent state. From Fig.~\ref{fig:2}(a), it is clear that  the combination of only non-Markovian channels does not always contribute to the speed-up of quantum evolution. It depends on the nature of the coupling strength and initial states. The behaviour of QSL time with respect to the memory is depicted in Fig.~\ref{fig:2}(b), and the coherence-mixedness impact on quantum speed limit time is shown in Fig.~\ref{fig:2}(c). In Fig.~\ref{fig:all_oun_rtn}, $\tau_{QSL}$ vs $\kappa\tau$ is plotted for the dynamics which involve only emission processes (nMAD) as well as the maps with both the emission and dephasing. In these cases, the emission process is always non-Markovian in nature, whereas the dephasing occurs in both (non)-Markovian regimes. From Fig.~\ref{fig:all_oun_rtn}, it is seen that considering dephasing noises (RTN and OUN) along with nMAD can  decrease the speed of evolution. The decoherence process due to the spontaneous emission and pure dephasing processes cannot always increase the speed of evolution compared with the decoherence process due to only spontaneous emission. In fact, for all range of parameter values, except for a small interval (Fig.~\ref{fig:all_oun_rtn}), adding dephasing channels decreases the speed of evolution for the initial states considered.

\begin{figure}[!htb]
    \includegraphics[width=\linewidth]{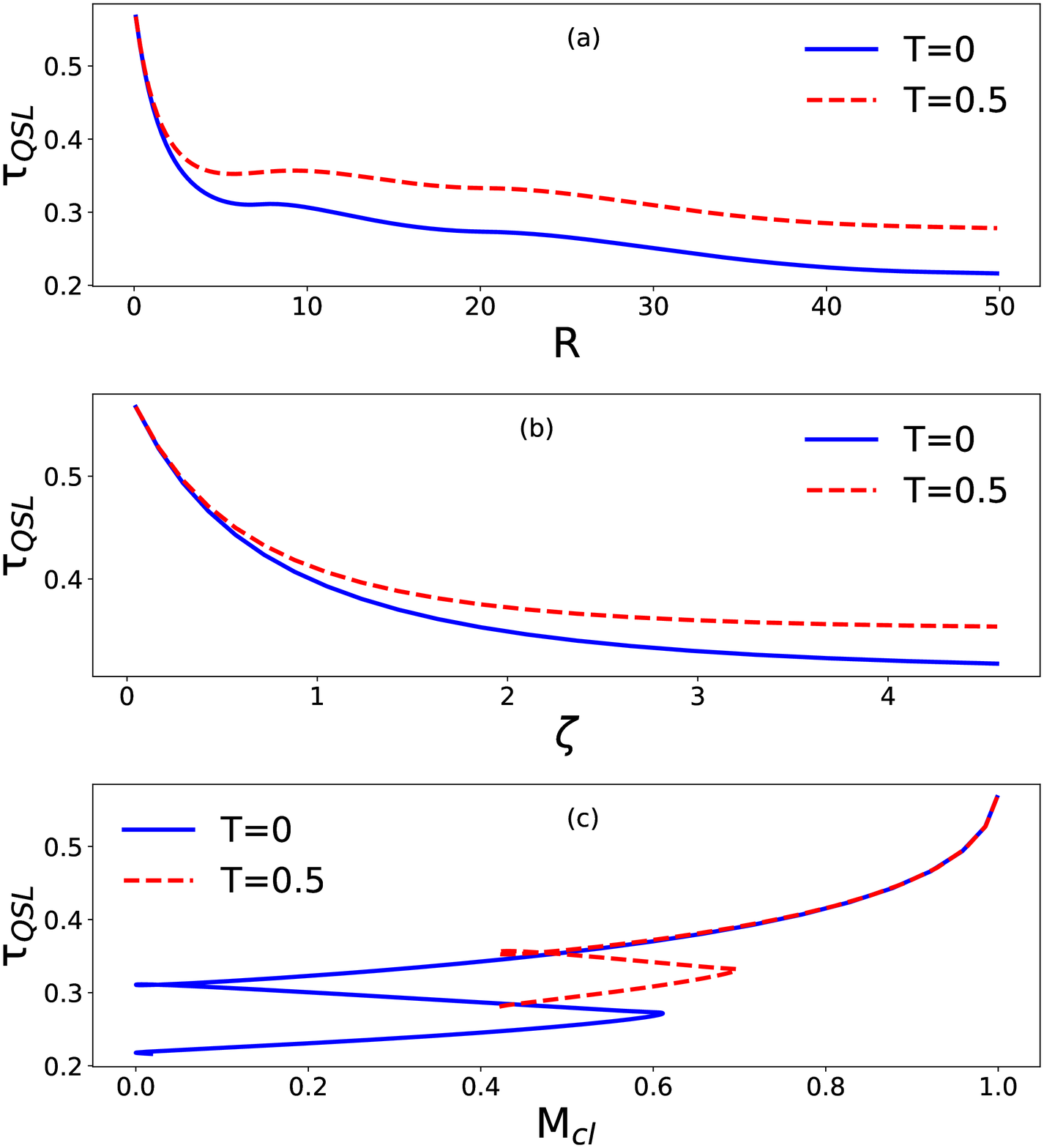}
    \caption{ ${\tau_{QSL}}$ vs. R, $\tau_{QSL}$ vs $\zeta$ and  ${\tau_{QSL}}$ vs $M_{cl}$ are  plotted  in subplots (a), (b) and (c), respectively, at zero and finite temperatures. QSL time for the pure state (${\frac{1}{\sqrt{2}}(|0\rangle+|1\rangle)}$) is estimated for temperature-dependent decay and dephasing processes at temperatures ${T=0,0.5}$ for super Ohmic spectral density s=4. We have c(0)=1, ${\omega_{c}=1}$, ${\upsilon=1}$ and the actual driving time ${\tau=1}$}\label{fig:3}
\end{figure}
\subsection{\label{sec:3.c}A phenomenological model}
We next  consider a phenomenological model which discusses the decoherence of states  due to the physical processes such as absorption, emission and pure dephasing. All the three rates ${\gamma_1(t),\gamma_2(t)}$ and ${\gamma_3(t)}$ take  nonzero values (Eq.~\ref{eq:phase_cov}) and reproduces the Markovian master equation in the appropriate limit ~\cite{PhysRevA.93.052103}. The decay rates are given as 
\begin{align}
    \gamma_1(t)&=2\mathcal{N}f(t) \label{eqn:g1},\\
    \gamma_2(t)&=2(\mathcal{N}+1)f(t)\label{eqn:g2},
\end{align}
where
\begin{align}
    f(t)&=-2\text{Re}\Big\{\frac{1}{c(t)}\frac{d c(t)}{dt}\Big\}, \\
    c(t)&=c(0)e^{-t/2}\Big[\text{cosh}(\sqrt{1-2R}t/2)+\frac{\text{sinh}(\sqrt{1-2R}t/2)}{\sqrt{1-2R}}\Big].
\end{align}
Here $R$ is a dimensionless constant greater than zero, which depicts the coupling between the system and the environment as well as the environmental spectral properties. ${\mathcal{N}(T)=[exp(\nu_0/T)-1]^{-1}}$ is the mean number of excitation in  modes of the thermal environment, where $T$ is the temperature and ${\nu_0}$ is the Bohr frequency. When $R < 1/2$, ${\gamma_1(t)}$ and ${\gamma_2(t)}$ are always positive. They become negative for certain time intervals when $R > 1/2$ and the dynamics become non-Markovian. At temperature $T=0$, ${\mathcal{N}(T)}=0$, and it reduces to non-Markovian amplitude damping. This matches with ${\gamma_2(t)}$ in Eq.~(\ref{eq:12}) with $l$ equal to one. 
The pure dephasing rate is given by
\begin{equation}
    \gamma_3(t)=2\int d\omega J(\omega) \text{coth}(\omega/k_B T)\text{sin}(\omega t),
    \label{eqn:spec}
\end{equation}
where the spectral density ${J(\omega)}$ is
\begin{equation}
    J(\omega)=\frac{\upsilon\omega^s}{\omega^s_c}e^{-\omega/\omega_c},
\end{equation}
with $\omega_c$ being the cut-off frequency and $\upsilon$ being a dimensionless constant. In the pure dephasing case, i.e., when $\gamma_1(t)=\gamma_2(t)=0$, the Ohmic parameter $s$ as a function of temperature determines the Markovianity of the model. The system is non-Markovian when $s > {s_{crit}(T)}$. The critical value of $s$ has a minimum of $s_{crit}(0)=2$ at $T=0$ and a maximum of $s_{crit}(T\rightarrow\infty)=3$ in the high-temperature limit. Using the time-independent and dependent generators in Eq.~\ref{tss_measure}, we calculate the channel's memory, $\zeta>0$ implies the presence of the quantum memory.  

With the maximally coherent state taken as the initial state, Fig.~\ref{fig:3}(a) shows the behaviour of ${\tau_{QSL}}$ with respect to $R$. We find that ${\tau_{QSL}}$ is higher at finite temperatures and continues to increase with the increase in temperature.  The increase in temperature tends to slow down the process. The change of QSL time for the memory $\zeta$ is depicted in  Fig.~\ref{fig:3}(b). Figure~\ref{fig:3}(c) shows the trajectories taken by the ${\tau_{QSL}}$ with respect to ${M_{cl}}$ at different temperatures. The system never reaches a pure state at finite temperatures corresponding to ${M_{cl}=0}$. 

\begin{figure}[h]
    \includegraphics[width=\linewidth]{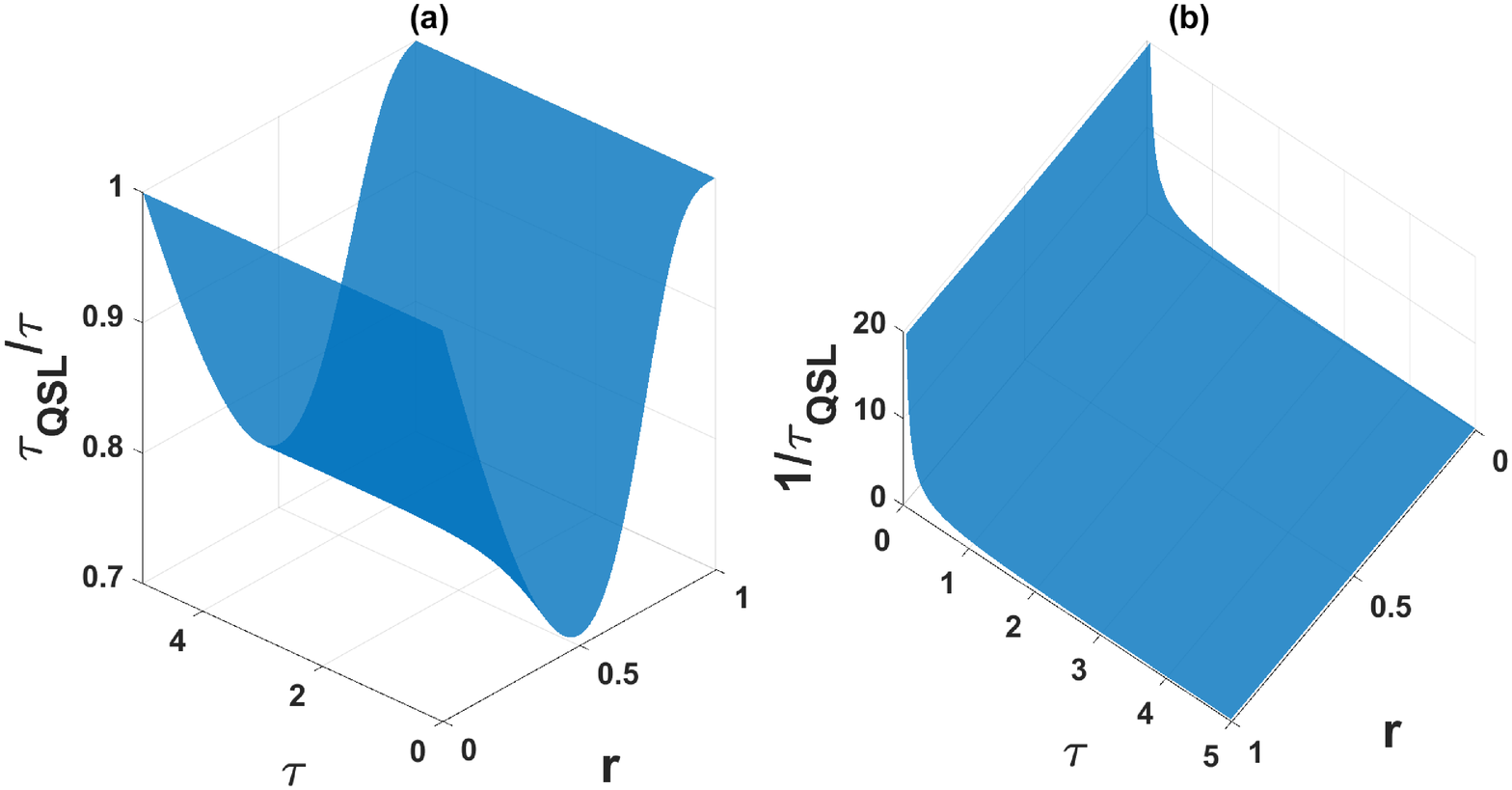}
    \bigbreak
    \includegraphics[width=\linewidth]{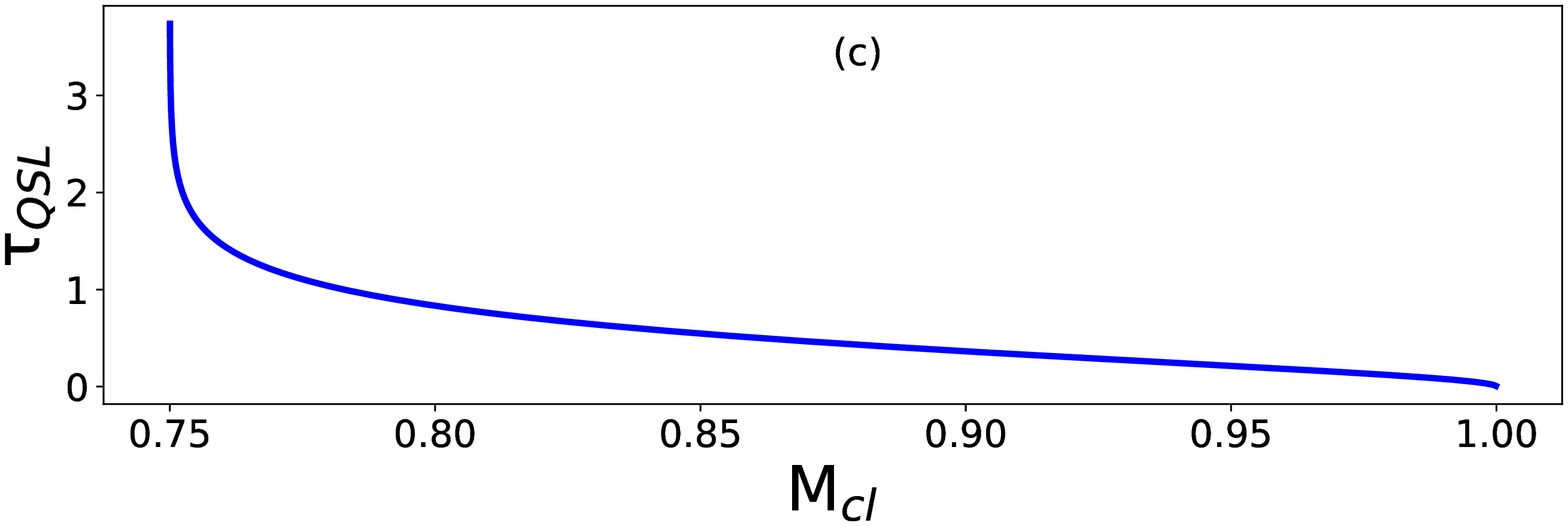}
    \caption{(a) ${\tau_{QSL}/\tau}$ and (b)1/${\tau_{QSL}}$ with respect to ${\tau}$ and r, (c) ${\tau_{QSL}}$ plotted as a function of ${M_{Cl}}$ with the maximally coherent state taken as the initial state for eternal CP-indivisible dynamics. Here $b=0.5$, $\nu=1$.}\label{fig:et_ncp}
\end{figure}

\subsection{\label{sec:3.d}Eternally CP-indivisible dynamics}
A family of non-unital eternal CP-indivisible dynamical maps was introduced in ~\cite{RussianFillipov} with the decay rates of absorption, emission and dephasing 
\begin{align}
    \gamma_1(t)&=2\nu(1+b),\\
    \gamma_2(t)&=2\nu(1-b),\\
    \gamma_3(t)&=-\frac{\nu(1-b^2)\text{sinh}(2\nu t)}{[1+b^2+(1-b^2)\text{cosh}(2\nu t)]},
\end{align}
Here, ${\nu>0}$. When ${|b|<1}$, ${\gamma_3(t)<0}$ for all t${>}$0, i.e., eternally indivisible dynamics which is neither unital nor commutative. \\
The SSS measure for this channel is $ \zeta=\frac{1}{T}\ln\left(\frac{1+b^2+(1-b^2)\text{cosh}(2\nu T)}{2}\right).$
As $\zeta>0$ for $T>0$, we see that this channel possesses memory.
In Fig.~\ref{fig:et_ncp}(a), we plot ${\tau_{QSL}}$ as a function of ${\tau}$, the driving time and $r$, where $r$ is used to parameterize the single qubit wave function, ${|\psi\rangle=\sqrt{r}|0\rangle+\sqrt{1-r}|1\rangle}$.
Depending on the initial state chosen, the behaviour of ${\tau_{QSL}}$ varies. We find ${\tau_{QSL}/\tau = 1}$ for two different states when ${r=0}$ and when ${r=1} $, while for all other states, it remains below 1. The upper bound to the Holevo rate decreases rapidly with the increase in driving time for all states, as can be seen in Fig.~\ref{fig:et_ncp}(b). The ${\tau_{QSL}}$ is plotted as a function of ${M_{cl}}$ in Fig.~\ref{fig:et_ncp}(c). With the maximally coherent state taken as the initial state, ${M_{cl}}$ first decreases from 1 as ${\tau_{QSL}}$ increases and then remains steady at 0.75 in Fig.~\ref{fig:et_ncp}(c). This can be explained using Eq.~(\ref{eq:trade-off}). In the limit $\tau\rightarrow\infty$, we find $M_{cl}=(1-b^2)$ which is equal to 0.75 for our chosen ${b=0.5}$.


\section{\label{sec:level4}Initial mixed states}
\begin{figure}[!htb]
    \includegraphics[width=\linewidth]{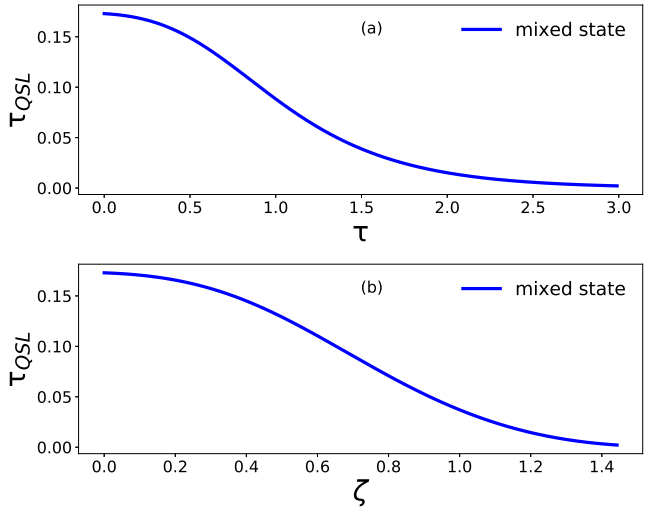}
    \caption{${\tau_{QSL}}$ vs. $\tau$ and ${\tau_{QSL}}$ vs. $\zeta$ are plotted  in subplots (a) and (b), respectively, for the eternally CP-indivisible channel. The initial mixed state has purity (tr${(\rho^2)}$) of 0.56 with ${r_x=r_y=r_z=0.2}$. Here $\tau_d=2$, $b=0.5$ and $\nu=1$.}\label{fig:fl_sec2}
\end{figure}

So far, we have considered only initial pure states. This section calculates the quantum speed limit time for initial mixed states under the eternal CP-indivisible dynamics. Different approaches are  used to  estimate the QSL time according to the purity of initial states~\cite{sun2015quantum,wu2018quantum,WuYu} The single qubit density matrix is
\begin{equation}
    \rho=\frac{1}{2}
    \begin{pmatrix}
    1+r_z && r_x-i r_y \\
    r_x+i r_y && 1-r_z 
    \end{pmatrix},
\end{equation}
where ${|r|\leq 1}$ with the inequality for mixed states.
The time-evolved density matrix for the eternally CP-indivisible channel then becomes,
\begin{widetext}
\begin{equation}
    \rho(t)=\frac{1}{2}
    \begin{bmatrix}
     1+e^{-2\nu t}(r_z+b(e^{2\nu t}-1))& (r_x-ir_y)e^{-2vt}\sqrt{\frac{1+b^2+(1-b^2)\text{cosh}(2\nu t)}{2}} \\
    (r_x+ir_y)e^{-2vt}\sqrt{\frac{1+b^2+(1-b^2)\text{cosh}(2\nu t)}{2}} & 1-e^{-2\nu t}(r_z+b(e^{2\nu t}-1))
    \end{bmatrix}.
\end{equation}
\end{widetext}
We use the modified ${\tau_{QSL}}$ with the factor ${\left(1+\sqrt{\frac{1-\text{tr}\rho_\tau^2}{1-\text{tr}\rho_t^2}}\right)}$ multiplied to Eq.~(\ref{B_spdlmt_1}). We look for the minimal time required for the system to evolve from a mixed initial state ${\rho_\tau}$ to the final state ${\rho_{\tau+\tau_d}}$. As the open system quantum evolution is non-unitary, ${\rho_{\tau+\tau_d}}$ is also a mixed state. The modified ${\tau_{QSL}}$~\cite{WuYu} is
\begin{equation}
    \tau_{QSL}=\frac{\text{sin}^2[\mathcal{B}_{\tau,\tau+\tau_d}]}{\frac{1}{\tau_d}\int_\tau^{\tau+\tau_d}dt \vert\vert \mathcal{L}(\rho_t)\vert\vert_{\textrm{op}}\Bigg(1+\sqrt{\frac{1-\textrm{tr}\rho_\tau^2}{1-\textrm{tr}\rho_t^2}}\Bigg)},
\end{equation}
where ${\mathcal{B}_{\tau,\tau+\tau_d}=\text{arccos}(\mathcal{F}(\rho_\tau,\rho_{\tau+\tau_d}))}$ and ${\mathcal{F}}$ is the super fidelity. Using this definition, ${\tau_{QSL}}$ is plotted with respect to $\tau$ and memory $\zeta$ in Figs.~\ref{fig:fl_sec2}(a) and ~\ref{fig:fl_sec2}(b), respectively.
They indicate that the minimum time required to achieve a particular overlap between two mixed states decreases with evolution.

\section{Action quantum speed limit}

Until now, we have used quantum speed limit time based on a geometrical approach. The  Geometrical approach for quantum speed limits relies on the idea that the geodesic distance between any two points is the shortest possible length connecting them. The speed of evolution depends on the path of evolution. Geometric quantum speed limits are not sensitive to instantaneous speed. This problem is resolved by calculating the action $\tau_{QSL}$~\cite{eoin}. Here, we optimize our action $\tau_{QSL}$. We implement this on the phase covariant map with $\gamma_{1}(t)=\gamma$, $\gamma_{2}(t)=\Gamma$, $\gamma_{3}(t)=0$. This is the generalized amplitude damping channel (system in contact with a thermal bath at non-zero temperature). We consider a pure initial state, $|\psi\rangle= \text{cos}(\frac{\theta}{2})|0\rangle+\text{sin}(\frac{\theta}{2})|1\rangle$. \\
Under the metrics chosen in this work, the action $\tau^a_{QSL}$
can be calculated as
\begin{equation}
    \tau^a_{QSL}=\frac{(\text{sin}^2[\mathcal{B}(\rho_0, \rho_t)])^2}{a^{\gamma}_{g}}.
\end{equation}
Here, $a^{\gamma}_{g}$ is the action which needs to be minimized along the path and is expressed as
\begin{align}
\nonumber
    a^{\gamma}_{g}&=\int_0^{\tau} dt \mathscr{L}(\rho_t, \dot{\rho}_t, \dot{q}(t))\\\nonumber
    &=\int^{\tau}_{0} dt ||\dot{\rho}(t)||^2_{\text{op}} \\
    &=\int_{0}^{\tau} dt \  \dot{q}(t)^2\left(\frac{\text{sin}^2 2\theta}{16(1-q(t))}+(\text{sin}^2 \theta -\eta)^2\right),
    \label{action}
\end{align}
\begin{figure}[!htb]
    \includegraphics[width=\linewidth]{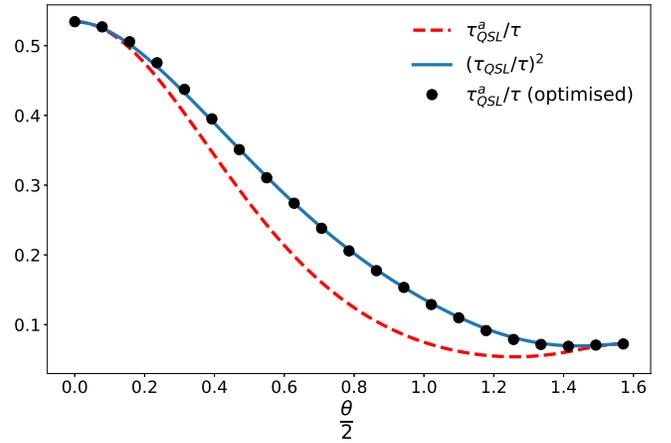}
    \caption{The red dotted line corresponds to the unoptimized action $\tau^a_{QSL}/\tau$ while the solid blue line corresponds to geometric $(\tau_{QSL}/\tau)^2$. The initial guess for $q(t)=t/\tau$. The black dotted line is the  result of  optimization of the path traversed, and we have $\tau=1$.}\label{fig:action}
\end{figure}
where $q(t)$ describes the path traversed by the channel and makes its appearance in the Kraus operators of the channel under consideration. $\mathscr{L}(\rho_t, \dot{\rho}_t, \dot{q}(t))$ is the control Lagrangian. Also, $\eta=\frac{1}{2}(1+\text{tanh}\beta)$, with $\beta=\frac{1}{2}\text{ln}\frac{\gamma}{\Gamma}$. The Hamiltonian is obtained using the Legendre transform. Modified gradient descent algorithm~\cite{Deffner_2014} is used to obtain the consecutive $\dot{q}(t)$'s. \\
From Fig.~\ref{fig:action}, we see that optimization of the  action leads to saturation of the Cauchy Schwartz inequality. Thus the geometric $\tau_{QSL}$ is seen to be an upper bound of the action $\tau^a_{QSL}$.

\section{\label{sec:level5}Conclusion}
As is known, the phase-covariant map describes the physical process involving absorption, emission and pure dephasing. This provides a convenient platform to study both (non)-unital processes from a common perspective. We made use of this to make an exhaustive study of QSL time for different sets of physical processes ((non)-unital, (non)-Markovian) within the framework of the phase covariant channel. For this, we considered maps that do not involve the absorption process and maps containing all three processes. For absorption-free phase covariant maps,  we considered the cases where both CP-(in)divisible (non)-Markovian dephasing are involved. For the initial maximal coherent state, it was found that the presence of a CP-indivisible non-Markovian dephasing map does not always speed up the quantum evolution.



In the phenomenological model, where absorption, emission and dephasing rates are considered together, the temperature was found to affect ${\tau_{QSL}}$ significantly, with ${\tau_{QSL}}$ increasing with the increase in temperature for the chosen initial state. Another feature of this study is the impact of the coherence-mixedness trade-off on ${\tau_{QSL}}$ for evolution generated by the phase-covariant map. Various features of the dynamics, such as information backflow, and temperature effects, could be ascertained from this. In general, the speed of evolution depends on the initial state and the channel parameters. A newly developed formulation for quantum speed limit called action speed limit, which takes into account the path of the evolution, was also considered and compared with the geometrical QSL time.
The generality of the present study implies that special cases of the phase-covariant map, such as the amplitude damping and the pure dephasing channels, can be easily obtained as simplifications of the models studied. 

\section*{Acknowledgement}
SB acknowledges support from the Interdisciplinary Cyber-Physical Systems (ICPS) programme of
the Department of Science and Technology (DST), India, Grant No.: DST/ICPS/QuST/Theme-1/2019/6. SB also acknowledges support from the Interdisciplinary Program (IDRP) on Quantum Information and Computation (QIC) at IIT Jodhpur.

\bibliography{apssamp}

\appendix
\section{\label{sec:levelA}}
The ${l_1}$ norm of coherence is given by:
\begin{equation} \label{eq:coh}
    C_{l_1}=\sum_{i\neq j} |\rho_{ij} |.
\end{equation}
For a qubit, this is the sum of the absolute value of the off-diagonal elements.
The mixedness, based on normalized linear entropy for a single qubit, is:
\begin{equation}\label{eq:mix}
    M_l(\rho)=2(1-Tr(\rho^2)).
\end{equation}
Using (\ref{eq:den}), (\ref{eq:coh}) and (\ref{eq:mix}), we find for the phase covariant noise:
\begin{align}
    C_{l_1}&=2|\alpha(t)|, \\
    M_l({\rho})&=4(p_1(t)-p_1(t)^2-|\alpha(t)|^2).
\end{align}
The trade-off between mixedness and coherence (${C_{l_1}^2+M_l({\rho}):=M_{cl}}$) is then calculated as:
\begin{equation}
    M_{cl}=4p_1(t)(1-p_1(t)) \leq 1.
\end{equation}

\end{document}